\documentclass[conference]{IEEEtran}
\usepackage[T1]{fontenc}
\usepackage{mathtools}
\usepackage{amssymb, amsthm, amsmath}
\usepackage{color}
\usepackage{graphicx}
\usepackage{epstopdf}
\usepackage{multirow}
\usepackage{braket}

\newcommand{\alis}{\hspace{-10pt}& &\hspace{-10pt}}

\usepackage{color}
\definecolor{Blue}{rgb}{0,0,1}
\definecolor{Red}{rgb}{1,0,0}

\title{A Quantum Optimization Algorithm for Single Machine Total Weighted Tardiness Minimization}

\author{\IEEEauthorblockN{Youhao S. Wang\\}
\IEEEauthorblockA{Union County Magnet High School \\
Scotch Plains, New Jersey, USA 07076 \\
Email: ywang1051@ucvts.org} \and

\IEEEauthorblockN{Julian Cheng\\}
\IEEEauthorblockA{School of Engineering \\
The University of British Columbia \\
Kelowna, British Columbia, Canada \\
Email: julian.cheng@ubc.ca}
}

\begin{document}

\maketitle

\begin{abstract}
A single machine total weighted tardiness minimization (TWTM) problem in operational planning is considered. The problem is formulated as an NP-hard constrained combinatorial problem, which has no known deterministic polynomial complexity solution using classical computing. Based on efficient Grover’s quantum search and Trugenberger’s quantum optimization algorithms, a novel efficient quantum optimization algorithm is proposed to solve the NP-hard single machine TWTM problem, which makes the desired solution satisfying the searching constraints and showing the minimal TWT value be measured with the highest probability. 
\end{abstract}


\section{Introduction}
Since quantum computers were proposed in the 1980's \cite{Benioff1980, Feynman1986}, quantum computing has attracted widespread interest as it appears to be especially more powerful than classical computing for certain types of problems \cite{PolakQIntro2000}. Some examples of powerful quantum algorithms include Shor's quantum factorization algorithm that finds factors of integers in polynomial time \cite{Shor1994, Shor1997} and Grover's quantum search algorithm, which is an optimal unstructured search algorithm for quantum computers  \cite{GroverSearch1996}. 

Grover's search algorithm uses quantum mechanics principles to search an unstructured list, in which items are arranged in a completely random manner and no knowledge about the structure of the solution is assumed or used. The algorithm identifies the item in the list satisfying a given condition as the solution. For the unstructured search problem, the computational complexity of classical algorithms grows at least at the order of the list size. However, the computational complexity of Grover's quantum search algorithm only grows at most at the order of the square root of the list size. For this type of problem, quantum computing is more efficient than classical computing. 

Furthermore, another class of search problems in which quantum computing excels is called the combinatorial search problem or combinatorial optimization problem. In these problems, a cost value is associated with each item in the searching list and the goal is to find the item associated with the minimum (or maximum) cost value. This type of problem is NP-hard and has no known classical solution that has computational complexity increasing in a polynomial relationship to the searching list size. While multiple suboptimal algorithms have been developed based on classical computers, none have achieved polynomial computational complexity. In \cite{QoptTrugenberger2002}, Trugenberger's quantum optimization algorithm was proposed for unconstrained combinatorial search problems based on quantum mechanics principles. Like Grover's quantum search algorithm, it manipulates quantum parallelism so that the desired solution can be measured with higher probability compared with nonsolutions. The computational complexity of this quantum optimization algorithm is independent of the list size. 

Combinatorial optimization problems with constraints occur in certain practical applications. For example, the total weighted tardiness minimization (TWTM) problem, which is a well-known NP-hard problem, can be found in operational planning. This problem requires construction of a schedule for a single machine with a fixed start time and multiple tasks with various due times that minimizes the sum of weighted tardiness of tasks relative to their respective due times \cite{PavlovBook2019}. The problem can be formulated as a constrained combinatorial optimization problem.  

To solve the TWTM problem, we propose a novel efficient quantum optimization algorithm based on Grover's quantum search algorithm and Trugenberger's quantum optimization algorithm to ensure the desired solution that satisfies the searching constraints and shows the minimal TWT value in the searching list will be measured with the highest probability. In the proposed quantum optimization algorithm, we also propose a novel cost function normalization method that makes the proposed algorithm more practical and powerful. The computational complexity of the proposed quantum optimization algorithm is thus determined by that of Grover's quantum search algorithm, which grows at most at the order of the square root of the searching list size. 

%
%
%
%

The rest of this paper is organized as follows. In Section II, the TWTM problem is first presented and then formulated as an NP-hard constrained binary combinatorial problem. The basic concept of quantum computing is reviewed briefly in Section III before presenting Grover's quantum search and Trugenberger's quantum optimization algorithms in Section IV. The proposed novel quantum optimization algorithm for the TWTM problem is presented in Section V. Finally, conclusions are drawn in Section VI.

\section{Problem Statement and Formulation} 
\label{section_formulation}
In this paper, we consider the TWTM problem on a single machine \cite{PavlovBook2019}. It is assumed that there is a single machine with a fixed start time and $M$ independent tasks each consisting of a single operation and with different job lengths $t_m$ ($m=1,\ldots,M$) and deadlines $d_m$ ($m=1,\ldots,M$). 

If the task schedule is determined as ${\mathcal S} = \{ i_1, i_2, \ldots, i_M \}$ with $i_k \in \{ 1, \ldots, M \}$ ($k=1,\ldots,M$), then the completion time of task $m$ is
\begin{equation}
C_m = \sum_{k=1}^{m} t_{i_k} ~~~~m=1,\ldots,M. 
\end{equation}
Therefore, the tardiness of task $m$ relative to its deadline is 
\begin{equation}
f_m = C_m - d_m ~~~~m=1,\ldots,M.
\end{equation}
The TWTM problem is to find a schedule ${\mathcal S}$ for the tasks that minimizes the sum of weighted tardiness of tasks relative to their deadlines, which can be expressed as
\begin{equation}
\min f = \sum_{m=1}^{M} w_m f_m, \label{TWT_cost}
\end{equation}
where $w_m$ is the weight of task $m$. Generally, greater weights are assigned to more important or urgent tasks. The problem is NP-hard in the strong sense and intractable. There does not exist any solution with deterministic polynomial time complexity in classical computing for the problem \cite{PavlovBook2019}. 

Equivalently, the TWTM problem can also be considered as how to schedule tasks into $M$ time slots whose durations are flexible so that the TWTM goal in (\ref{TWT_cost}) is achieved. Accordingly, we define a task-scheduling vector ${\mathbf x}$ as
\begin{equation}
{\mathbf x} = \left[ ({\mathbf x}_M^1)^T ~({\mathbf x}_M^2)^T ~\cdots ~({\mathbf x}_M^M)^T\right]^T \in \{0,1\}^{M^2 \times 1}, \label{Def_x_vector}
\end{equation}
where ${\mathbf x}_M^m \in \{0,1\}^{M \times 1}$ ($m=1,\ldots,M$) shows how the $m$th time slot is scheduled. Due to the fact that only one task can be scheduled in the $m$th time slot, the vector ${\mathbf x}_M^m$ has only $M$ possible values, i.e., ${\mathbf x}_M^m \in \left\{ [1 ~0 ~\ldots ~0]^T, ~[0 ~1 ~\ldots ~0]^T, ~\cdots, ~[0 ~0 ~\ldots ~1]^T \right\}$. The position of $1$ in it indicates which task is allocated in time slot $m$. 

Based on the definition of ${\mathbf x}$, the TWTM problem can be re-formulated as a constrained binary combinatorial problem as
\newcounter{MYtempeqncnt}
\setcounter{MYtempeqncnt}{\value{equation}}
\renewcommand{\theequation}{\arabic{equation}\alph{abc}}
\newcounter{abc}
\setcounter{abc}{1}
\begin{eqnarray}
\underset{\mathbf{x}}{\arg} ~\mathrm{min} ~f(\mathbf{x}) \alis = \sum_{m=1}^{M} w_m \cdot \left[\left( {\mathbf P}_M^m {\mathbf x} \right)^T {\mathbf t} - d_m\right]  \label{TWTObj}\\
\setcounter{equation}{\value{MYtempeqncnt}+1} \stepcounter{abc}
\textup{subject to: } \alis \left[ {\mathbf I}_M ~{\mathbf I}_M ~\cdots ~{\mathbf I}_M \right] \cdot {\mathbf x} = {\mathbf 1}_M \label{TWTcons_exclusive}  \\ 
\setcounter{equation}{\value{MYtempeqncnt}+1} \stepcounter{abc}
\alis {\mathbf x} \in \{0,1\}^{M^2 \times 1}, \label{TWTcons_xint}
\end{eqnarray}
\renewcommand{\theequation}{\arabic{equation}}
\newcounter{MYtempeqncnt_xt}
\setcounter{MYtempeqncnt_xt}{\value{equation}}
\hspace{-0.15in} where 
\begin{equation}
{\mathbf P}_M^m = \left[ \underbrace{{\mathbf I}_M \cdots {\mathbf I}_M}_{m}~\underbrace{{\mathbf 0}_{MM} \cdots {\mathbf 0}_{MM}}_{M-m} \right]
\end{equation}
and ${\mathbf t} = [t_1 ~t_2 \cdots ~t_M]^T$. In the above formulation, the notations ${\mathbf I}_M$ and ${\mathbf 0}_{MM}$ represent identity and zero matrices of size $M \times M$, respectively and ${\mathbf 1}_M$ represents column vector composed of $M$ $1$'s. The constraint in (\ref{TWTcons_exclusive}) ensures that all tasks are scheduled into different time slots without any conflict, which will be referred to as $h({\mathbf x})$ in the following discussion.

\section{Basics of Quantum Computing}

This section will not provide a thorough introduction to general quantum mechanics or quantum computing. Instead, the focus will be on explaining only the concepts of quantum computing needed in the algorithms to be covered \cite{PolakQIntro2000}.

\subsection{Quantum State Space and Bra/Ket Notation} \label{Subsec_qstate}
For quantum computing, we only need to consider finite quantum systems. Skipping the details of physics associated with the state spaces of finite quantum systems, we can describe quantum state spaces and their transformations using vectors and matrices in finite dimensional complex vector spaces, and bra/ket notation is an equivalent and more convenient notation to use. Kets, in the form of $\Ket{x}$, are used to denote column vectors, while bras, in the form of $\Bra{x}$, denote the conjugate transposes of $\Ket{x}$. For example, the two-dimensional orthonormal bases $\{[1 ~0]^T, [0 ~1]^T\}$ can also be conveniently represented as $\{ \Ket{0}, \Ket{1} \}$.

\subsection{Superposition and Measurement}
Based on Section \ref{Subsec_qstate}, the state of a quantum bit, or qubit, can be represented by a vector in a two-dimensional complex vector space with orthonormal bases $\Ket{0}$ and $\Ket{1}$. A qubit may be in a superposition state $a \Ket{0} + b \Ket{1}$, where $a$ and $b$ are complex numbers such that $|a|^2 + |b|^2 = 1$. When the qubit is measured, the probability that the measurement result is state $\Ket{0}$ is $|a|^2$, and the probability that the measurement result is state $\Ket{1}$ is $|b|^2$. 

To help us understand superposition states, we can think of a coin. The coin heads up could represent the qubit being in one of the two orthogonal basis states, for example $\Ket{0}$, and the coin tails up could represent the qubit being in the other orthogonal state $\Ket{1}$. Based on this analogy, the qubit in superposition would be the state the coin is in when it is flipped and still spinning. The qubit still incorporates both of its orthogonal states, but is not existing in any of them. 

Superposition is fragile. A superposition state will only last until the qubit is measured, which occurs whenever a qubit in superposition is put through any process or device that makes its state known. Upon being measured, the state of the qubit will collapse from the superposition state into one of the two orthogonal basis states associated with the measurement process or device. Continuing with the coin analogy, the measurement leads to the coin falling on one of its two sides. The likelihood that the qubit will be measured in one orthogonal state as opposed to the other could be different if $|a|^2 \neq |b|^2$.

\subsection{Multiple Qubits}
The state space of two qubits, each with bases $\{ \Ket{0}, \Ket{1} \}$ is spanned by the four-dimensional orthonormal bases $\{[1 ~0 ~0 ~0]^T, [0 ~1 ~0 ~0]^T, [0 ~0 ~1 ~0]^T, [0 ~0 ~0 ~1]^T\}$, which can be more conveniently represented as $\{ \Ket{00}, \Ket{01}, \Ket{10}, \Ket{11} \}$. Note that $\Ket{00} = \Ket{0} \otimes \Ket{0}$, $\Ket{01} = \Ket{0} \otimes \Ket{1}$, $\Ket{10} = \Ket{1} \otimes \Ket{0}$, and $\Ket{11} = \Ket{1} \otimes \Ket{1}$, with $\otimes$ representing the tensor product \cite{TensorProductWebpage}. In the more general case of $n$ qubits, the state space has $2^n$ orthonormal bases. The state of $n$ qubits can thus be represented by a vector in this $2^n$-dimensional complex vector space. 

\subsection{Quantum Transformation and Quantum Gate} \label{Subsec_qgate}
Any linear transformation on a complex vector space can be described by a matrix. A matrix is unitary if the product of itself and its conjugate transpose is an identity matrix and therefore all unitary matrices must have inverses and describe reversible transformations. Any legitimate quantum transformation can be described by such a unitary matrix and therefore is a unitary transformation of a quantum state space. Therefore, legitimate quantum transformations are reversible. It has also been shown that all classical computations can be reverted. Thus all classical computations can be implemented by using legitimate quantum transformations. Basic quantum transformations are implemented as quantum gates, which form the building blocks of quantum computers. As an example, the Hadamard gate (H-gate) is defined by the transformation $\Ket{0} \rightarrow \frac{1}{\sqrt{2}} \left(\Ket{0} + \Ket{1}\right)$ and $\Ket{1} \rightarrow \frac{1}{\sqrt{2}} \left(\Ket{0} - \Ket{1}\right)$ with the unitary transformation matrix
\begin{equation}
\frac{1}{\sqrt{2}} \left[ \begin{array}{cc}
1 & 1 \\ 
1 & -1 
\end{array}\right]. 
\end{equation}

\subsection{Parallelism}
From Section \ref{Subsec_qgate}, it is known that any classical function $g$ with $i$ input bits and $j$ output bits can be implemented on a quantum computer. Therefore, it can be shown that the transformation 
\begin{equation}
U_g: \Ket{x, y} \rightarrow \Ket{x, y \oplus g(x)} \label{Def_Ug}
\end{equation}
is unitary for any function $g$. In (\ref{Def_Ug}), the notation $\oplus$ denotes the bitwise exclusive OR (xor) operation. By defining $y = 0$, any classical function $g(x)$ can be computed using the linear transformation $U_g$ defined in (\ref{Def_Ug}). 

In addition, since $U_g$ is linear, if it is applied to an input in a superposition state, it is equivalently applied to all basis states simultaneously. Thus, we can compute $f(x)$ for all possible bases inputs to obtain the corresponding outputs in one step. This is called quantum parallelism.    

\section{Quantum Searching and Quantum Combinatorial Optimization} \label{section_quantum_algs}

\subsection{Quantum Search Algorithm} \label{GroverSearchSubsec}
Based on quantum mechanics, Grover proposed an unstructured search algorithm \cite{GroverSearch1996}, in which no knowledge about the structure of the solution is assumed or used. It is assumed that the size of the searching set is no greater than $2^n$ and any item in the searching set can be represented by a combination of $n$ binary bits. There is only one item in the searching set that satisfies a given condition and needs to be identified as the searching solution. The detailed steps of the Grover's quantum search algorithm are listed below \cite{PolakQIntro2000, GroverSearch1996}. 

\begin{enumerate}

\item Initialize an $n$-qubit message and an additional control qubit, so that all qubits in the message are in state $\Ket{0}$ and the control qubit is in state $\Ket{1}$.

\item Pass each of the $n$ qubits in the message and the control qubit through H-gates. Thus, at the outputs of the H-gates, the superposition states of the message qubits ($\Ket{\psi}$) and the control qubit ($\Ket{b}$) are
\begin{eqnarray}
\Ket{\psi} \alis = \frac{1}{\sqrt{2^n}} \sum_{i=0}^{2^n-1} \Ket{x_i} \nonumber \\
\alis = \frac{1}{\sqrt{2^n}} \left( \Ket{0 \cdots 0 0} + \Ket{0 \cdots 0 1} + \cdots \nonumber \right. \\
\alis ~~~~~~~~~~~~~~~~\left. + \Ket{1 \cdots 1 1} \right) \\
\alis \hspace{-0.15in} \textup{and} \nonumber \\
\Ket{b} \alis = \frac{1}{\sqrt{2}} \left( \Ket{0} - \Ket{1} \right),
\end{eqnarray}
respectively. 

\item Continue to pass the message and the control qubits through the gatearray $\textup{U}_H$ with unitary transformation
\begin{equation}
U_H: \Ket{x_i, b} \rightarrow \Ket{x_i, b \oplus H(x_i)}, \label{Def_UH}
\end{equation}
which is defined in the same way as in (\ref{Def_Ug}). In (\ref{Def_UH}), $H$ is the classical function representing the searching condition, with $H(x_i) = 0$ if $x_i$ does not satisfy the condition and $H(x_i) = 1$ if $x_i$ satisfies the condition. At the output of the gatearray $\textup{U}_H$, the $n$ left most qubits correspond to the message, whose superposition state is denoted by $\Ket{\psi^{\prime}}$.

\item Apply the unitary transformation 
\begin{equation}
\mathbf{A} = \left[ \begin{array}{cccc}
\frac{2}{2^n}-1 & \frac{2}{2^n} & \cdots & \frac{2}{2^n} \\ 
\frac{2}{2^n} & \frac{2}{2^n}-1 & \cdots & \frac{2}{2^n} \\
\vdots \\
\frac{2}{2^n} & \frac{2}{2^n} & \cdots & \frac{2}{2^n}-1 
\end{array}\right] \in {\mathcal R}^{2^n \times 2^n} \label{Def_A}
\end{equation}
to the $n$ qubits corresponding to the message. The superposition state of the resulting message is denoted by $\Ket{\psi^{\prime \prime}}$.

\item Repeat steps 3 and 4 up to $\left\lfloor \frac{\pi}{4} \sqrt{2^n} \right\rfloor$ times \cite{QSearchBounds1998}.

\item Measure the state of the message, which represents the searching solution. 

\end{enumerate}

In step 3, it is shown that 
\begin{equation}
U_H(\Ket{\psi,b}) = \Ket{\psi^{\prime},b}
\end{equation}
and that the amplitude of the quantum state satisfying the searching condition is converted to its negative value, while the amplitudes of the other states are kept the same. That is, in the first round when steps 3 and 4 are executed, we have
\begin{equation}
\Ket{\psi_1^{\prime}} = \frac{1}{\sqrt{2^n}} \left( \sum_{i \in {\mathcal S}_0} \Ket{x_i} - \sum_{i \in {\mathcal S}_1} \Ket{x_i} \right), \label{psi_1_pri}
\end{equation}
with 
\begin{equation}
\begin{array} {c}
{\mathcal S}_0 = \left\{ i | H(x_i) = 0, ~i = 0, \ldots, 2^n-1 \right\}~ \\
{\mathcal S}_1 = \left\{ i | H(x_i) = 1, ~i = 0, \ldots, 2^n-1 \right\}.
\end{array}
\end{equation}
Under the assumption of only having a single searching solution, the set size of ${\mathcal S}_1$ is one, i.e., we have $\left|{\mathcal S}_1 \right| = 1$ and $\left|{\mathcal S}_0 \right| = 2^n-1$. Note that in (\ref{psi_1_pri}), a subscript is added in the notation $\Ket{\psi^{\prime}}$ to indicate the number of the round.

In step 4, the amplitudes of the superposition state $\psi^{\prime}$ is inverted relative to twice of their average value. It is shown that in the first round when steps 3 and 4 are executed, the average amplitude is
\begin{equation}
A_1 = \frac{1}{\sqrt{2^n}} - \frac{1}{\sqrt{2^{3n-2}}}, \label{A_1}
\end{equation} 
and thus
\begin{eqnarray}
\Ket{\psi_1^{\prime\prime}} = \alis a_{10}  \sum_{i \in {\mathcal S}_0} \Ket{x_i} + a_{11}  \sum_{i \in {\mathcal S}_1} \Ket{x_i} \\
= \alis \left( \frac{1}{\sqrt{2^n}} - \frac{2}{\sqrt{2^{3n-2}}} \right)  \sum_{i \in {\mathcal S}_0} \Ket{x_i} \nonumber\\
 + \alis \left( \frac{3}{\sqrt{2^n}} - \frac{2}{\sqrt{2^{3n-2}}} \right)  \sum_{i \in {\mathcal S}_1} \Ket{x_i}. \label{psi_1_pripri}
\end{eqnarray}
Similar to (\ref{psi_1_pri}), in (\ref{A_1}) and (\ref{psi_1_pripri}), subscripts are added in $A$ and $\Ket{\psi^{\prime\prime}}$ to indicate the number of the round. It is easily seen that $a_{11} > a_{10}$. After finishing $r$ ($1 \leq r \leq \left\lfloor \frac{\pi}{4} \sqrt{2^n} \right\rfloor$) rounds, at the output of step 4, the superposition state of the message is denoted by
\begin{equation}
\Ket{\psi_r^{\prime\prime}} = a_{r0}  \sum_{i \in {\mathcal S}_0} \Ket{x_i} + a_{r1}  \sum_{i \in {\mathcal S}_1} \Ket{x_i}, \label{psi_k_pripri}
\end{equation}
with $a_{r1} > a_{r0}$. Ideally, the number of rounds $r$ is chosen such that $a_{r1} \gg a_{r0}$.


\subsection{Quantum Combinatorial Optimization Algorithm}
A quantum optimization algorithm is proposed by Trugenberger for unconstrained combinatorial optimization problems \cite{QoptTrugenberger2002}. In the proposed algorithm, the solution of the problem, which minimizes (or maximizes) a cost function $F(x)$, with $x$ being the problem variable represented by a combination of $n$ binary bits, is found based on quantum computation. In the following discussion, the function $F_n(x)$ denotes the normalized cost, whose value is obtained from $F(x)$ and falls between $0$ and $+1$.  The problem solution is denoted by $x^{\ast}$. The detailed steps of the Trugenberger's quantum optimization algorithm are listed below \cite{PolakQIntro2000} \cite{QoptTrugenberger2002}. 

\begin{enumerate}

\item Initialize an $n$-qubit variable register, so that all qubits in the register are in state $\Ket{0}$. Pass each of the $n$ variable qubits through H-gates.

\item Initialize a $c$-qubit control register, so that all qubits in the register are in state $\Ket{0}$. Initialize the index of the control qubit being considered as $k=1$. 

\item Pass the $k$th control qubit through an H-gate.  

\item Continue to pass the combined variable and control qubits through the controlled gate $\textup{U}_{c}^{\pm}$, which applies the unitary transformation
\begin{equation}
\textup{diag}\left\{ U_{c}^{\pm}(x_0), U_{c}^{\pm}(x_1), \cdots, U_{c}^{\pm}(x_{2^n-1}) \right\}. \label{controlledUpnTrans}
\end{equation}
If the control qubit being examined, which is $k$th in the control bit register, is in state $\Ket{0}$, then
\begin{equation}
U_{c}^{\pm}(x_i) = e^{j \frac{\pi}{2} F_{n}(x_i)} ~~~~i=0,\ldots,2^n-1 \label{Up_trans}
\end{equation}
where $j$ is the imaginary unit satisfying $j^2 = -1$ and the normalized cost $F_n(x)$ is calculated based on $F(x)$ as
\begin{equation}
F_n(x) = \frac{F(x) - F_{\min}}{F_{\max} - F_{\min}}, \label{Fn_T}
\end{equation}
where $F_{\min}$ and $F_{\max}$ are the strict lower and upper bounds of the cost function $F(x)$, respectively, and thus $0 \leq F_n(x) \leq 1$. If the control qubit being considered is in state $\Ket{1}$, then
\begin{equation}
U_{c}^{\pm}(x_i) = e^{-j \frac{\pi}{2} F_{n}(x_i)} ~~~~i=0,\ldots,2^n-1. \label{Un_trans}
\end{equation}

\item Pass the $k$th control qubit through an H-gate again. 

\item If $k < c$, then update the value of $k$ as $k+1$ and go back to step 3. Otherwise, continue with step 7.

\item Measure the control register. If all $0$'s (or $1$'s) are obtained in the measurement result for a minimization (or maximization) problem, then continue with step 8. Otherwise, re-start from step 1.

\item Measure the variable register to get the solution $x^{\ast}$.

\end{enumerate}

In the first iteration when $k=1$, after step 3, the superposition state of the combined variable and control qubits is
\begin{eqnarray}
\Ket{\varphi_{1,1}} = \alis \frac{1}{\sqrt{2^{n+1}}} \sum_{i=0}^{2^n-1} \Ket{x_i, \underbrace{0 0 \cdots 0}_{c \textup{ bits}}} \nonumber \\
\alis + \frac{1}{\sqrt{2^{n+1}}} \sum_{i=0}^{2^n-1} \Ket{x_i,\underbrace{1 0 \cdots 0}_{c \textup{ bits}}},
\end{eqnarray}
where $\Ket{x_i}$ ($i=0,\ldots,2^n-1$) represents the states of the $n$-qubit variable, which may take the values of $\underbrace{\Ket{0 \cdots 0 0}}_{n \textup{ bits}}$, $\underbrace{\Ket{0 \cdots 0 1}}_{n \textup{ bits}}$, $\cdots$, or $\underbrace{\Ket{1 \cdots 1 1}}_{n \textup{ bits}}$.

After step 5, it is shown that the superposition state of the combined variable and control then becomes  
\begin{eqnarray}
\Ket{\varphi_{1,2}} = \alis \frac{1}{\sqrt{2^{n}}} \sum_{i=0}^{2^n-1} \cos \Delta_i \cdot \Ket{x_i, \underbrace{0 0 \cdots 0}_{c \textup{ bits}}} \nonumber \\
\alis + j \frac{1}{\sqrt{2^{n}}} \sum_{i=0}^{2^n-1} \sin \Delta_i \cdot \Ket{x_i, \underbrace{1 0 \cdots 0}_{c \textup{ bits}}},
\end{eqnarray}
where $\Delta_i$ is defined as $\frac{\pi}{2} F_{n}(x_i)$.

After all $c$ iterations are finished and before the control register is measured, the superposition state of the combined variable and control is shown to be
\begin{eqnarray}
\Ket{\varphi_{c,2}} = \frac{1}{\sqrt{2^{n}}} \sum_{k=0}^{2^c-1} j^{|y_k|} \sum_{i=0}^{2^n-1} \alis \cos^{c-|y_k|} \left( \Delta_i \right) \cdot \sin^{|y_k|} \left( \Delta_i \right) \nonumber \\
\alis \cdot \Ket{x_i, y_k},
\end{eqnarray}
where $\Ket{y_k}$ ($k=0,\ldots,2^c-1$) and $|y_k|$ represent the states of the $c$-qubit control and the number of $1$'s in $y_k$, respectively. 


\section{A Quantum Optimization Algorithm for TWTM}
It is known that the TWTM problem is NP-hard without any known exact polynomial time solution using classical computers. Motivated by the constrained binary combinatorial problem formulation of the TWTM problem in (\arabic{MYtempeqncnt_xt}), and Grover's quantum search algorithm and Trugenberger's quantum optimization algorithm presented in Section \ref{section_quantum_algs}, we propose a quantum optimization algorithm for TWTM problems. 

Based on the definition of ${\mathbf x}$ in (\ref{Def_x_vector}), the number of qubits needed to represent the variables in the TWTM problem is $N = M \log_2 M$, where $\log_2 M$ is an integer. In case the original problem does not satisfy this condition, we can add some dummy tasks with zero job lengths and infinite deadlines until this condition is satisfied. In addition, it is also noted that the strict lower and upper bounds of the cost function in (\ref{Fn_T}) may not be available conveniently in practical scenarios, which results in difficulty in the implementation of the proposed algorithm. Therefore, we propose a cost function normalization method as
\begin{equation}
F_n(x) = \frac{1}{1+e^{-\beta \left( F(x)-\alpha \right)}}, \label{Fn_prop}
\end{equation}
where $\beta$ is a large positive real number and $\alpha$ can be chosen as the mid-value between two arbitrary cost function values which, in general, can be obtained conveniently in most practical situations. 

The detailed steps of the proposed algorithm are listed below. 

%

\begin{enumerate}

\item Initialize an $N$-qubit variable and an additional control qubit, so that all variable qubits are in state $\Ket{0}$ and the control qubit is in state $\Ket{1}$.

\item Pass each of the $N$ variable qubits and the control qubit through Hadamard gates (H-gates).  

\item Continue to pass the variable and the control qubits through the gatearray $\textup{U}_h$ defined in (\ref{Def_UH}), with the searching condition $h$ defined in (\ref{TWTcons_exclusive}). 

\item Apply the unitary transformation ${\mathbf A}$ defined in (\ref{Def_A}) to the $N$ qubits corresponding to the variable. 

\item Repeat steps 3 and 4 up to $\left\lfloor \frac{\pi}{4} \sqrt{\frac{2^N}{M!}} \right\rfloor$ times.

\item Discard the control qubit. 

\item Initialize another control qubit so that the control qubit is in state $\Ket{0}$ and pass the control qubit through an H-gate.

\item Pass the current combined variable and control qubits through the controlled gate $\textup{U}_{c}^{\pm}$ defined in (\ref{controlledUpnTrans}), (\ref{Up_trans}), and (\ref{Un_trans}), where the normalized cost function in (\ref{Fn_prop}) with the cost function in (\ref{TWTObj}) is used. 

\item Pass the control qubit through an H-gate again. 

\item Measure the control qubit. If a $0$ is obtained in the measurement result, then continue with step \ref{LastStep}. Otherwise, restart from step 1.

\item Measure the variable qubits to obtain the solution $x^{\ast}$. \label{LastStep}

\end{enumerate}


It is shown that in the first round when steps 3 and 4 of the proposed algorithm are executed, the superposition state of the variable after step 4 is given by
\begin{eqnarray}
\Ket{(\psi_1^{TWT})^{\prime\prime}} = \alis \left( \frac{1}{\sqrt{2^N}} - \frac{2M!}{\sqrt{2^{3N-2}}} \right)  \sum_{i \in {\mathcal S}_0} \Ket{x_i} \nonumber\\
 + \alis \left( \frac{3}{\sqrt{2^N}} - \frac{2M!}{\sqrt{2^{3N-2}}} \right)  \sum_{i \in {\mathcal S}_1} \Ket{x_i}, 
\end{eqnarray}
with $\left|{\mathcal S}_1 \right| = M!$ and $\left|{\mathcal S}_0 \right| = 2^N-M!$.

It is also shown that after step 9 of the proposed algorithm and before the control qubit is measured, the superposition state of the combined variable and control is
\begin{eqnarray}
\alis \Ket{\Phi^{TWT}} \nonumber \\
\alis =  a_{r0} \sum_{i \in {\mathcal S}_0} \cos  \Delta_i  \cdot \Ket{x_i, 0}  + a_{r1} \sum_{i \in {\mathcal S}_1} \cos  \Delta_i  \cdot \Ket{x_i, 0}  \nonumber \\
\alis + j a_{r0} \sum_{i \in {\mathcal S}_0} \sin  \Delta_i  \cdot \Ket{x_i, 1} + j a_{r1} \sum_{i \in {\mathcal S}_1} \sin \Delta_i  \cdot \Ket{x_i, 1}. ~~~~~~ \label{SuperState_TWT}
\end{eqnarray}
Since $a_{r1} > a_{r0}$ and cosine is a decreasing function on the interval between $0$ and $2 \pi$, it can be seen from (\ref{SuperState_TWT}) that the term corresponding to the desired solution that satisfies the searching constraints and shows the minimal TWT value has the largest amplitude $a_{r1} \cdot \cos \Delta_{i,\min}$, with $\Delta_{i,\min}$ denoting the minimum among all values of $\Delta_{i}$. Therefore, based on the superposition measurement rule in quantum mechanics, the probability that the measurement results of the variable qubits show the desired solution $x^{\ast}$, which equals $| a_{r1} \cdot \cos \Delta_{i,\min} |^2$, will be the highest. 


\section{Conclusion}
In this paper, we have formulated the single machine TWTM problem as an NP-hard constrained combinatorial problem. A novel efficient quantum optimization algorithm has been proposed to solve the NP-hard single machine TWTM problem based on Grover's quantum search and Trugenberger's quantum optimization algorithms in literature. Alongside the proposed quantum optimization algorithm, a more powerful cost function normalization method has also been proposed. 

Due to the limitations of quantum computing platforms currently available, the proposed quantum optimization algorithm has not been implemented, as have other algorithms proposed in this field in literature. As future work, we will investigate the implementation of the proposed algorithm on available quantum computing platforms.

\end{document}